\documentclass{mem}
\usepackage{natbib}\usepackage{txfonts}\usepackage{balance}
\usepackage{graphicx}
\usepackage[a4paper,breaklinks,dvipdfm]{hyperref}
\idline{75}{282}
\begin{document}
\def\teff{$T\rm_{eff }$}
\def\kms{$\mathrm {km s}^{-1}$}

\title{
Compact Massive Objects in Galaxies
}

   \subtitle{}

\author{
I. \,Tosta e Melo 
\and R. \, Capuzzo-Dolcetta
          }

\institute{
Sapienza Universit\`{a} di Roma, P.le A. Moro 5, I-00185 Roma, Italy.
\email{iara.tosta.melo@gmail.com}
}

\authorrunning{Iara }

\titlerunning{CMOs in Galaxies}

\abstract{
The central regions of galaxies show the presence of super massive black holes and/or very dense stellar clusters. Both objects seem to follow similar host-galaxy correlations, suggesting that they are members of the same family of Compact Massive Objects. We investigate here a huge data collection of Compact Massive Objects properties to correlate them with absolute magnitude, velocity dispersion and mass of their host galaxies. 
\keywords{galaxies, nuclei, nuclear star clusters, and supermassive black holes. }
}
\maketitle{}

\section{Introduction}

Various studies suggest that massive galaxies, both elliptical and spiral, harbor a SuperMassive Black Hole (SMBH) in their centers, with masses between $10^6 - 10^9$ M$_\odot$. Galaxies across the entire Hubble sequence also show the presence of massive and compact stellar clusters referred to Nuclear Star Clusters (NSCs). In elliptical galaxies, the NSCs are also referred to as Resolved Stellar Nuclei (RSN). Despite their different morphologies, some galaxies of the Local Group present both a NSC and a SMBH, and in such case, the SMBH is surrounded by the NSC. Those objects follow a similar host-galaxy correlation, suggesting that they are members of the same family of Compact Massive Objects (CMOs). CMOs constitute a interplay between either SMBH or a compact stellar structure (NSCs, nuclear stellar Disks or resolved stellar nuclei). Studies in the literature also showed that the NSC mass versus the host galaxy velocity dispersion ($\sigma$) relation is roughly the same observed for SMBHs. Graham (2012) claimed, instead, that the $M_{NSC}-\sigma$ relation is shallower for NSCs ($M_{NSC} \propto \sigma^{1.5}$) than for SMBHs. Here we investigate these scaling correlations for a set of NSC and SMBH data wider than what already studied in the literature

\section{The Database and Method}

Turner et. al (2012) selected 43 galaxies of the Fornax Cluster with early-type morphologies, using the Hubble Space Telescope (HST) Advanced Camera of Survey (ACS). Côté et al. (2006) analysed the nuclei of a sample of 100 early-type galaxies in the Virgo Cluster. Georgiev $\&$ Böker (2014) presented the properties of 228 nuclear stellar cluster in nearby late-type disk galaxies observed with the WFPC2/Hubble Space Telescope (HST/WFPC2). den Brok et al. (2014) analysed the light profile of 200 early-type dwarf galaxies using the HST/ACS Coma Cluster Survey. We also included in our sample 89 galaxies having a dynamical detection of the central black hole mass.

Our first aim was to estimate CMOs mass for each of the catalogues presented before, and compare such values with the various observed and derived parameters of their host galaxies. The masses of the stellar nuclei were calculated using the stellar mass-to-light ratio versus color index (CI) relation given by Bell et al. (2003):

\begin{equation}
\log_{\rm 10}(M/L)=a_{\lambda} + b_{\lambda}CI,
\label{eq.mass-to}
\end{equation}

The galaxy masses were obtained by means of the virial theorem:

\begin{equation}
M_g=\frac{\beta R_{\rm eff} \sigma^{2}}{G},
\label{eq.virial}
\end{equation}

where $R_{\rm eff}$ is the galaxy effective radius, and $ \beta=5 $, as given in Ferrarese et al. (2006). 
All the values for SMBH database were taken from the literature.

\section{Results}

The existence of evident correlations indicates a direct link among large galactic spatial scales and the much smaller scales of nuclear environment. We have fitted scaling correlations connecting the CMOs mass to various properties of their host galaxies: $M_{\textrm{B}}$, $\sigma$, and galaxy mass. To obtain the final value of the two fit parameters, $a$ and $b$, for each galaxy cluster, we applied a linear best fit, i.e. least $\chi^2$ method. 

NSCs are more common in fainter galaxies, with magnitudes between $-21\leq M_{\textrm{B}} \leq-13$, than in brighter galaxies as shown in Fig. 1(a), and Galaxies with larger masses have more massive CMOs.
The presence of NSCs is much more frequent in galaxies with lower velocity dispersion, Fig. 1(b).
 
\begin{figure}[]
\resizebox{\hsize}{!}{\includegraphics[clip=true]{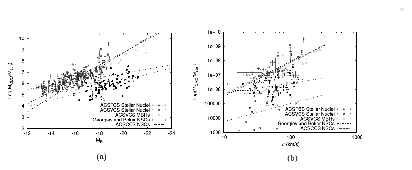}}
\caption{
\footnotesize
CMOs masses versus $M_{\textrm{B}}$ (a),and  $\sigma$ of the host galaxies. The black dashed line, black dashed pointed line, black two pointed line, and black pointed line are the least square fit of the Fornax, Virgo, HST/WFCP2, and Coma sample, respectively.
}
\label{iara}
\end{figure}

\subsection {Summary}

Our findings are:
(i) galaxies brighter than $M_{\textrm{B}} =-18$ host SMBHs, and the existence of such objects in bright galaxies reconcile with the existence, in most of the cases, of an AGN. The lack of NSCs in faint galaxies may be related to the small number of globular clusters in galaxies; (ii) we reconfirmed that, with our updated set of data, as one moves to fainter galaxies, the nuclei become the dominant feature while MBHs tend to become less common and, perhaps, entirely disappear at the fainter end of the galaxy luminosity distribution; (iii) the $M_{NSC}-\sigma$ correlation shows a slope between 1.5 and 3, which is in good agreement with theoretical findings by Arca-Sedda and Capuzzo-Dolcetta (2014); (iv) as suggested by various theoretical-simulation scenarios, the dearth of NSCs in bright galaxies may be due to the presence of a single or binary SMBH which either evaporates the compact stellar structure or tidally destroy the infalling star clusters (Arca-Sedda et al. (2016)).

\begin{acknowledgements}
I. Tosta e Melo acknowledges CAPES-Brazil for support through the grant 9467/13-0.

\end{acknowledgements}

\bibliographystyle{aa}

\end{document}